\documentclass[]{article} 
\usepackage{emulateapj}
\usepackage{graphicx}
\usepackage{multirow}

\advance \voffset by -0.5cm\relax

\usepackage{color}

\def\msun{{\rm ~M}_{\odot}}
\def\rsun{{\rm ~R}_{\odot}}

\def\mpy{{\rm ~M}_{\odot} {\rm ~yr}^{-1}}

\begin{document}

\title{The fate of Cyg X-1: an empirical lower limit on BH-NS merger rate}

 \author{Krzysztof Belczynski\altaffilmark{1,2},
         Tomasz Bulik\altaffilmark{1,3}, Charles Bailyn\altaffilmark{4}}

 \affil{
     $^{1}$ Astronomical Observatory, University of Warsaw, Al.
            Ujazdowskie 4, 00-478 Warsaw, Poland\\
     $^{2}$ Center for Gravitational Wave Astronomy, University of Texas at
            Brownsville, Brownsville, TX 78520, USA\\
     $^{3}$ UMR ARTEMIS, CNRS, University of Nice Sophia-Antipolis, Observatoire 
            de la Cote d'Azur, BP 4229, 06304, Nice Cedex 4, France\\
     $^{4}$ Department of Astronomy, Yale University, P.O. Box 208101, New
            Haven, CT 06520, USA\\
 }
 
\begin{abstract}
The recent distance determination allowed precise estimation of the orbital
parameters of Cyg X-1, which contains a massive $14.8\msun$ BH with
a $19.2\msun$ O star companion. This system appears to be the clearest
example
of a potential progenitor of a BH-NS system.  We follow
the future evolution of Cyg X-1, and show that it will soon encounter a
Roche
lobe overflow episode, followed shortly by a Type Ib/c supernova and the
formation of a NS. It is demonstrated that in majority of cases
($\gtrsim70\%$) the supernova and associated natal kick disrupts
the binary due to the fact that the orbit expanded significantly in the
Roche lobe overflow episode.
In the reminder of cases ($\lesssim30\%$) the newly formed BH-NS
system is too wide to coalesce in the Hubble time. Only
sporadically ($\sim1\%$) a Cyg X-1 like binary may form a coalescing BH-NS
system given a favorable direction and magnitude of the natal kick. If Cyg
X-1 like channel (comparable mass BH-O star bright X-ray binary) is 
the only or dominant way to form BH-NS binaries in the Galaxy we can estimate the 
empirical BH-NS merger rate in the Galaxy at the level of $\sim 0.001$ Myr$^{-1}$.
This rate is so low that the detection of BH-NS systems in gravitational
radiation is highly unlikely, generating Advanced LIGO/VIRGO detection rates
at the level of only $\sim 1$ per century. If BH-NS inspirals are in fact
detected, it will indicate that the formation of these systems proceeds via
some alternative and yet unobserved channels.
\end{abstract}


\section{Introduction}

Estimates for the rates of gravitational radiation (GR) sources from
coalescing degenerate binaries are typically 
performed with population synthesis methods 
(e.g., Lipunov, Postnov \& Prokhorov 1997; Bethe \& Brown 1998; 
De Donder \& Vanbeveren 1998; Bloom, Sigurdsson \& Pols 1999; Fryer, Woosley 
\& Hartmann 1999; Nelemans, Yungelson \& Portegies Zwart 2001, Voss \& Tauris 
2003 or more recently Belczynski et al. 2010). 
These studies attempt to explain the past evolution of 
the given observed binary or stellar population and put some constraints on the
physics of stellar/binary evolution (e.g., Valsecchi et al. 2010).  
Nevertheless, there are often critical model parameters that are poorly
constrained.  

Recently, Bulik, Belczynski \& Prestwich (2011) have taken a different approach,
examining specific binary systems with well established parameters
and investigating their future evolution. If a binary
is chosen close to the end of its life (e.g., the formation of double
compact object) such a method has potentially great predictive power as many
unknowns relating to its prior binary and stellar evolution can be avoided.  
In particular, Bulik et al. (2011) considered two high mass X-ray
binaries (HMXBs), IC10 X-1 and NGC300 X-1, 
and showed that these systems will soon form close black hole + black hole
(BH-BH) systems that will merge within Hubble time and produce strong GR signature. 
This provided an empirical lower limit of detection chances for the current GR 
instruments without direct reference to population synthesis methods.

In this study, we consider the future evolution of one of the most
interesting binaries known in our Galaxy: Cyg X-1. Recently, the 
distance to this system was determined by radio parallax and other 
methods (Reid et al. 2011; Xiang et al. 2011), 
allowing the basic
parameters of this binary to be  firmly established (Orosz et al. 2011). This 
HMXB hosts one of the most massive ($15 \msun$) Galactic BHs in a close
orbit around a massive ($20 \msun$) O star. Since the companion is in the
mass range for neutron star formation, we have selected this system to 
investigate yet another unobserved population of potential GR sources: 
black holes + neutron star (BH-NS) systems.

\section{Estimates}

\subsection{The future evolution of Cyg X-1}

To evolve the system forward in time we use evolutionary prescriptions
incorporated in the {\tt StarTrack} population synthesis code (Belczynski et al.
2002). The evolution of the system is relatively simple and we do not
need any population synthesis tools at this point. We start off with the
best estimate of current binary parameters; the black hole with the mass 
$M_{\rm BH}=14.8 \msun$, the optical star with the mass $M_{\rm opt}=19.2 \msun$ 
and radius of $R_{\rm opt}=16.2 \rsun$ and the orbital period $P_{\rm orb}=5.6$d 
(Orosz et al. 2011). 

\begin{figure}
\includegraphics[width=1.0\columnwidth]{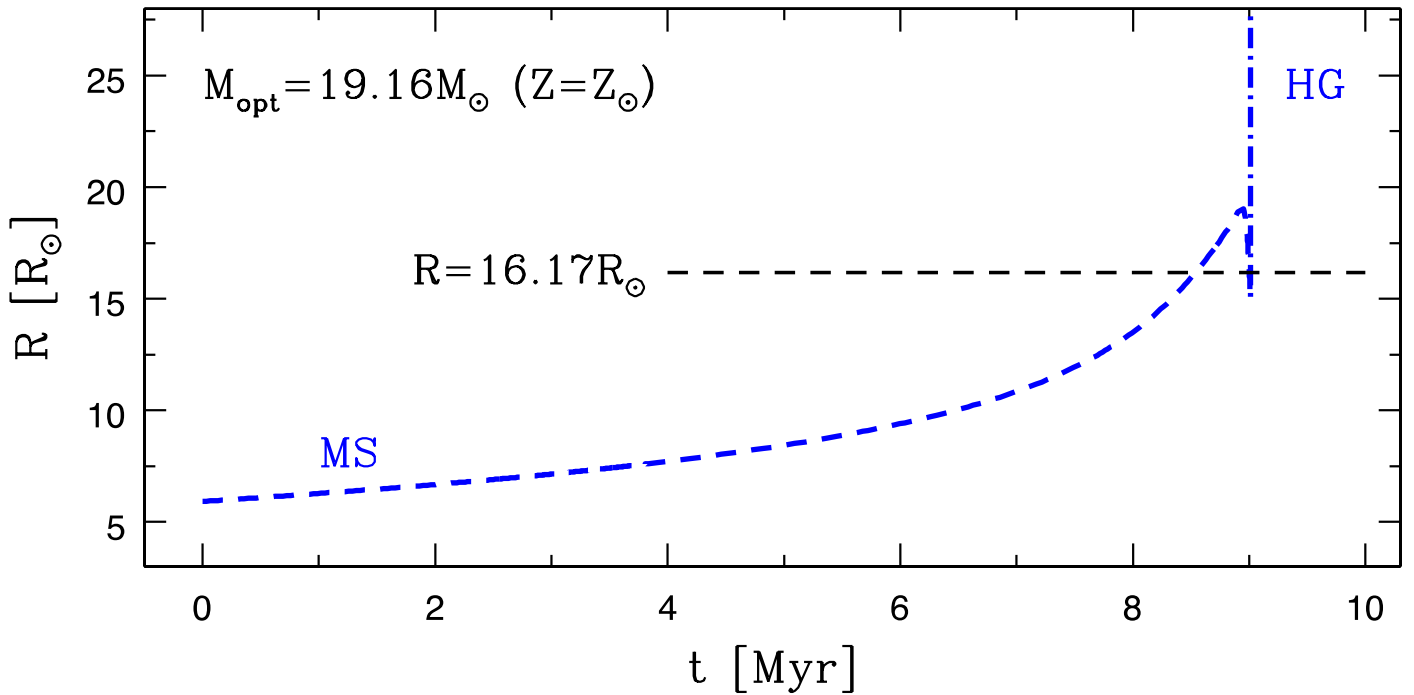}
\caption{Radius evolution of optical star in Cyg X-1. Current radius is
found at $R=16.17 R_\odot$ and that places the star at the end of its Main
Sequence (dashed line) or at the beginning of Hertzsprung gap (dot-dashed
line). Since the star is very close to its Roche lobe ($R_{\rm lobe}=17.24 R_\odot$ 
for the orbital period
$P_{\rm orb}=5.6$d) and since it has not yet started RLOF it means that the
star is on Main Sequence at the first intersection of its evolutionary
track and radius line of $R=16.17 R_\odot$. Once the star increases its
radius by about $1 R_\odot$ it will start RLOF while still on Main
Sequence and the RLOF will last through Hertzsprung gap (see Fig.~\ref{rlof}).
The radius evolution is taken from single star models of Hurley, Pols \& 
Tout (2000).} 
\label{radius}
\end{figure}

\begin{figure}
\includegraphics[width=0.7\columnwidth]{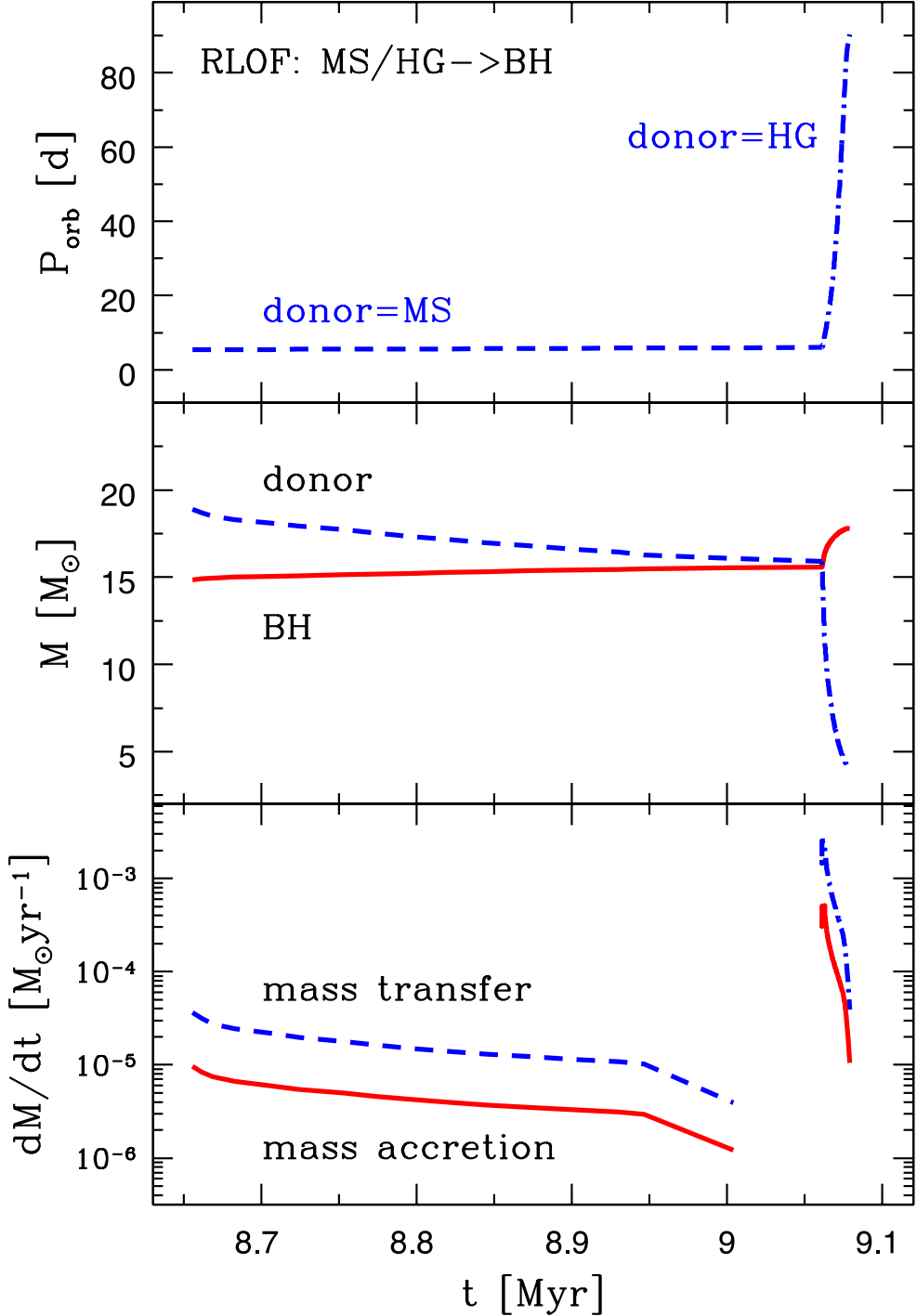}
\caption{Evolution of Cyg X-1 through RLOF that will start in about $10^5$ yr.
{\em Bottom panel:} Mass transfer rate from the massive donor star is very high. However,
it is much lower while the donor is on Main Sequence ($1-3 \times 10^{-5}
\mpy$: dashed line) as 
compared to the transfer during Hertzsprung gap ($10^{-3}-10^{-4} \mpy$:
dot-dashed line). Mass accretion
onto BH is factor of $\sim 3-5$ lower than the transfer rate to account for
the fact that BH cannot accept all the transferred material. Note that the 
accretion rate is significantly higher than typically employed Eddington rate 
($5 \times 10^{-7} \mpy$ for a $15 \msun$ non-spinning BH) as we account for 
more realistic super critical accretion (slim disk ADAF) onto a rapidly 
spinning BH with $a=0.9$ (Gou et al. 2011). 
{\em Middle panel:} The donor loses most of its mass to become a $4 \msun$
helium core with a small H envelope. Most of the donor mass ($\sim 9 \msun$) is 
lost from the system, while BH increases its mass from $14.8$ to $17.8 \msun$.
{\em Top panel:} Period of the system changes from currently observed 5.6
days to 90 days. System becomes wide after the RLOF due to the
non-conservative mass exchange and mass ratio reversal (most of the 
mass is accreted onto BH while the donor became the less massive component
of the binary).} 
\label{rlof}
\end{figure}

The optical companion is almost filling its Roche lobe and will start Roche
lobe overflow (RLOF) in less than 0.2 Myr while still on Main Sequence (see
Fig.~\ref{radius}). 
The mass ratio is close to unity so we do not expect the common envelope
evolution, but rather a stable RLOF phase (Belczynski et al. 2008b,
Wellstein, Langer \& Braun 2001). 
However, the mass transfer rate 
may reach quite high values while the donor is moving through Hertzsprung
gap (HG). The evolution of the system is presented in Figure~\ref{rlof}. 
Mass transfer rate is calculated using physical properties of the donor and
the system parameters (Belczynski et al. 2008a). The mass accretion onto BH
is calculated using the slim disk models
(e.g., Abramowicz et al. 1988; Ohsuga et al. 2005; Ohsuga 2007) and thus can
significantly exceed the classical Eddington limit (see Belczynski et al.
2008b for details). The BH increases its mass to $M_{\rm BH}=17.8 \msun$,
while the optical star loses most of its mass $M_{\rm opt}=4.2 \msun$ to
become a massive helium core with a bit of H-rich envelope. Note that the 
majority of the mass lost from the donor is lost from the system (highly
non-conservative case). The 
period of the system increases to $P_{\rm orb}=90$d. 
RLOF stops as the donor decreases in size due to the lost of its H-rich
envelope. The massive helium or Wolf-Rayet (WR) star, 
(($R_{\rm WR} \lesssim 1 \rsun$) is well 
within its Roche lobe $R_{\rm lobe} \approx 60 \rsun$. After some wind mass 
loss ($\sim 0.5 \msun$) and about $2$ Myr after the RLOF termination, the 
WR star explodes in Type Ib/c supernova and forms a neutron star (NS). 

There is about $2 \msun$ mass loss in the supernova and that is not enough to
disrupt the system. However, the pre-supernova binary is rather wide
($P_{\rm orb} = 104$d, and semi-major axis $a=250 \rsun$ due to
additional orbit expansion caused by the mass loss from the WR star), 
and any large natal kick tends to disrupt the binary. 
In Table~\ref{sn} we list the disruption and survival
probabilities for two assumptions about the distribution of
kick velocities. The ``full'' kicks are adopted from
the velocity distribution of the Galactic single pulsars that is best
described by a Maxwellian with 1D $\sigma=265$ km s$^{-1}$ (Hobbs et al. 2005)
and uniform distribution of orientation. 
As there may be some observational and theoretical evidence that natal kicks
are smaller in close binary systems (see the discussion in Belczynski et al.
2010) we also explore a ``half'' kick model with kicks drawn from the 
same distribution but with $\sigma=132.5$ km s$^{-1}$.  
As expected for such a wide system, the binary disruption and formation
of two single compact objects is most likely: 94\% and 74\% for full and half 
kicks, respectively. Less likely, but still quite probable is the formation
of a wide BH-NS system: 5\%  and 25\%. The least likely is the formation of the
close BH-NS system with the coalescence time below Hubble time ($13.47$
Gyr): 0.2\% and 0.8\%.
In order for the system to form a close BH-NS binary, the supernova explosion 
must produce a significant kick ($V_{\rm kick} \sim 150$ km s$^{-1}$) that
sends the NS toward the BH. 

Additionally, we have investigated an evolutionary scenario in which the
RLOF starts early in the HG. Since at this point the optical star has a more
massive core as compared with the above example, it will retain more mass
through RLOF (only the H-rich envelope is transferred/lost). Since less mass is
transferred/lost the orbital expansion is not as dramatic and the final 
orbital period of the binary at the supernova explosion is $P_{\rm orb} = 
61.7$d ($a=174 \rsun$). This obviously leads to higher survival
probabilities and higher chance of the close BH-NS formation: 
0.4\%  and 1.4\% for the full kick and half kick models, respectively.

\subsection{Rate Estimates}

Cyg X-1 has been detected because of its unusually strong X-ray
brightness. Let us consider the question of the formation rate of
Cyg X-1 like binaries in our Galaxy. The upper limit on the X-ray 
active phase in Cyg X-1 is set by the evolutionary time of the secondary,
which is $10$ Myr for a $20 \msun$ star. We employ this upper limit. 
Had we adopted more formal estimate of $5$ Myr (it takes about $5$ Myr for 
a massive star to form a BH that is already in the system and start an X-ray
phase) the estimated detection rate of BH-NS inspirals in 
gravitational radiation would increase by factor of 2. 
Given that we see just one such object and assuming that the current 
star formation rate is representative across the lifetime of the Galaxy we 
can infer the formation rate of Cyg X-1 like binaries as one per $10$ Myr, i.e. 
$r \approx 10^{-7}$ yr$^{-1}$. 
Our simulations show that the chance of forming
a merging BH-NS system is between  $2 \times 10^{-3}$ and $14 \times
10^{-3}$ (see bottom row of Tab.~\ref{sn}) from a Cyg X-1 like progenitor. These means
that the formation rate of BH-NS binaries from Cyg X-1 like progenitors
will be only  $2-14\times 10^{-10}$ yr$^{-1}$. This means that only
a handful ($2-14$) close BH-NS systems might have been formed over a
$10$ Gyr history of the Milky Way.
For comparison the Galactic empirical NS-NS merger rate is significantly
higher: NS-NS $3-190\times 10^{-6}$ yr$^{-1}$ (Kim, Kalogera \& Lorimer
2010).  

The implied Advanced LIGO/VIRGO detection rate follows once we determine 
the range of these detector for a BH-NS system. Given the final mass of the BH
($M_{\rm BH} = 17.8 \msun$) and assuming the mass of a NS to be 
canonical $1.4M \msun$, we obtain the chirp mass of the newly formed system 
to be $M_{\rm chirp} \equiv (M_{\rm BH} M_{\rm opt})^{3/5} (M_{\rm BH} + 
M_{\rm opt})^{-1/5} = 3.8 \msun$. Since the range for the Advanced detectors for NS-NS
system with a chirp mass of $1.2 \msun$ is $300$ Mpc, we obtain the range 
for such BH-NS systems of $786$ Mpc (the detection distance scales like
$\propto M_{\rm chirp}^{5/6}$). If we adopt the Milky Way like galaxies
density in the local Universe to be $0.01$~Mpc$^{-3}$ (e.g., O'Shaughnessy
et al.\ 2008), then within such distance there should be $2.0 \times 10^7$ 
galaxies. Combining it with the Galactic rate we obtain the
detection rate in the range $0.4-2.8 \times 10^{-2}$ yr$^{-1}$,
or a detection every 36 to 250 years in the Advanced LIGO/VIRGO.

The observational uncertainities on component masses ($\lesssim 2\msun$) do 
not play a significant role on our findings and qualitatively they do not 
change any of our results. The same applies to the stellar wind mass loss 
rate from the companion star as the $20\msun$ loses only about $1\msun$
during its main sequence (e.g., Vink, de Koter \& Lamers 2001). The major
uncertainities arise in the orbital evolution during RLOF in which 
$\sim 15\msun$ is lost/exchanged and supernova outcome (natal kicks).

\section{Discussion}

So far we have discussed only one binary, Cyg X-1, as a potential progenitor of 
BH-NS system. 
Tomsick \& Muterspaugh (2010) list several known Galactic HMXBs with a NS 
and a companion massive enough to potentialy produce a BH: Vela X-1 ($24\msun$), 
XTEJ1855-026 ($25\msun$), 4U1907+09 ($28\msun$) and GX301 ($40\msun$). 
The expansion of massive companions will eventually lead to a RLOF. These
systems due to extreme mass ratio will evolve into common envelope that will
result in a merger of a NS with its companion aborting the BH-NS formation. 
Such outcome follows from the fact that the mass of a NS is so small in
respect to the mass of the envelope of a companion that there is not enough
orbital energy to eject companion envelope (e.g., Webbink 1984). 
An extragalactic system LMC X-3 hosting a $10\msun$ BH was belived to have
a companion massive enough to form a NS. However, recent spectral analysis that 
takes into account irradiation of the companion, indicates that it is a B5
dwarf setting its mass at about $5\msun$ (Val-Baker, Norton \& Negueruela 2007) 
and makes it a WD progenitor.

HMXBs like Cyg X-1 are wind fed, and thus their non-degenerate stars do
not fill their Roche lobes. It is therefore curious that those binaries 
for which good system parameters are known tend to be close to RLOF. 
Cyg X-1 is an example: our analysis implies that the
system will begin RLOF in $0.2$ Myr, after a lifetime about 50 times 
longer than that.  An even more extreme case is that of LMC X-1, in which
an $11\msun $ black hole is found in a 3.9 day orbit with a $32\msun $
companion.  The companion currently fills over 90\% of its Roche lobe, 
implying that the system will undergo RLOF within $\approx 0.1$ Myr (Orosz
et al. 2009).
Note that the high mass of the companion implies that RLOF will result
in unstable mass transfer and the likely formation of a common envelope
system --- this LMC X-1 is unlikely to produce any sort of double degenerate
binary. 
There are only a few HMXBs with precisely known system parameters, so this
tendency to be alarmingly near RLOF may be a statistical anomaly.  
Alternatively, it may be a selection effect: a system in which the 
companion is far from RLOF will have a smaller fraction of its stellar wind
accrete onto the compact object, and thus a lower X-ray luminosity.  

Nevertheless, it may be worth entertaining the idea that this effect is neither
a statistical glitch nor an observational bias, but that there is some
unknown physical process that halts the growth of the companion star near
the Roche lobe boundary.  We note that the gravitational potential becomes
shallow near the $L_1$ point, and  X-ray irradiation becomes a bigger effect 
for companion stars that present a relatively large cross section. Both of
these effects may dramatically change the surface structure and stellar
winds of the companion star as it approaches the Roche lobe, conceivably in 
a self-limiting way.  Such winds might be observable through the strength,
shape and variability of emission lines associated with the wind.  If
such an effect is present,
the evolutionary path of the binary system may prove to be 
quite different from what is commonly assumed, and what we have assumed 
above.

To evaluate the possible effect this might have on rates of GR sources,
we consider how the future evolution of an HMXB like Cyg X-1 might proceed
if it never reaches RLOF.  As a limiting case, we constrain the orbital
period to remain at its currently observed value of 
$P_{\rm orb}=5.6$d. In fact, we expect the orbital period to increase
somewhat (even without RLOF) as mass is lost from the system in stellar
wind.  
In this case, the survival through the supernova and the formation of
the close BH-NS system is expected in $8.4\%$ of cases and corresponds to
Advanced LIGO/VIRGO detection rate of $\sim 1$ per decade. So despite the fact
that we have violated (in favor of producing close BH-NS systems) our
current
understanding of the stellar evolution we still do not get enough BH-NS
mergers to expect detection in gravitational waves. 

Thus we find that if indeed BH-NS mergers are observed as
GR sources, their immediate precursors will not be systems like Cyg X-1
that are currently observed.

\acknowledgements
Authors acknowledge the hospitality of the Aspen Center for Physics and
support from MSHE grants N N203 302835 (TB, KB); N203 404939, N N203 511238, 
NASA Grant NNX09AV06A to the UTB Center for Gravitational Wave Astronomy (KB)
and NSF-AST grant 0707627 (CB).

\clearpage

\begin{deluxetable}{lccc}
\tablewidth{350pt}
\tablecaption{BH-NS Formation  Statistics\tablenotemark{a}}
\tablehead{Fate/$P_{\rm orb}$\tablenotemark{b} & $104$d & $61.7$d & $5.6$d}

\startdata
SN disruption  & 0.944 (0.743) & 0.914 (0.673) & 0.680 (0.446)\\
Wide BH-NS     & 0.054 (0.249) & 0.082 (0.313) & 0.237 (0.500)\\
Close BH-NS    & 0.002 (0.008) & 0.004 (0.014) & 0.083 (0.054)\\
\enddata
\label{sn}
\tablenotetext{a}{
Fraction of Cyg X-1 like systems that after the second supernova will be 
disrupted, will form a wide BH-NS or close BH-NS (merger time shorter than 
the Hubble time). The fractions are given for the full natal kicks with 
$\sigma=265$ km s$^{-1}$ (or half kicks with $\sigma=132.5$ km s$^{-1}$). 
}
\tablenotetext{b}{
Numbers for $P_{\rm orb}=104$ and $61.7$d correspond to physical system modeling
with RLOF starting while the optical star is on MS and HG, respectively. 
The last model is unphysical, no RLOF was assumed and the orbital period was 
kept constant at $P_{\rm orb}=5.6$d through the evolution (see the Discussion). 
}
\label{sn}
\end{deluxetable}

\clearpage

\clearpage

\end{document}